\documentclass[preprint,showpacs,superscriptaddress,12pt]{revtex4-1}
\usepackage{amsmath,amssymb,graphicx}
\def\be{\begin{equation}}
\def\ee{\end{equation}}
\def\bea{\begin{eqnarray}}
\def\eea{\end{eqnarray}}
\def\bes{\begin{subequations}}
\def\ees{\end{subequations}}

\begin{document}
\title{Stern-Gerlach effect of multi-component ultraslow optical solitons
via electromagnetically induced transparency}
\author{Zhiming Chen}
\affiliation{State Key Laboratory of Precision Spectroscopy and
Department of Physics, East China Normal University, Shanghai
200062, China}

\author{Guoxiang Huang}
\email[Email: ]{gxhuang@phy.ecnu.edu.cn} \affiliation{State Key
Laboratory of Precision Spectroscopy and Department of Physics, East
China Normal University, Shanghai 200062, China}
\date{\today}

\begin{abstract}
We propose a scheme to exhibit a Stern-Gerlach effect of $n$-component ($n>2$) high-dimensional ultraslow optical solitons in a coherent atomic system with $(n+1)$-pod level configuration via electromagnetically induced transparency (EIT). Based on Maxwell-Bloch equations, we derive coupled (3+1)-dimensional nonlinear Schr\"{o}dinger equations governing the spatial-temporal evolution of $n$ probe-field envelopes. We show that under EIT condition significant deflections of the $n$ components of coupled ultraslow optical solitons can be achieved by using a Stern-Gerlach gradient magnetic field. The stability of the ultraslow optical solitons can be realized by an optical lattice potential contributed from a far-detuned laser field.
\end{abstract}

%\ocis{020.1670, 190.5530, 270.5530}

\maketitle

%%%%%%%%%%%%%%%%%%%%%%%%%%%%%%%%%%%%%%%%%%%%%%%%%%%%%%%%%%%%%%%%%%%%%%%%%%%%%%%%%

\section{Introduction}{\label{Sec:1}}

%%%%% Notice %%%%

% We should mention:
% 0. What is SG effect and its significance
% 1. This work is an analog of the space quantization analog of the SG effect of N-component atoms
% 2. This work is a generalization of Hang and Huang's work to multi-components
% 3. The long expressions should be put into Appendix
% 4. Point out the physical and math significance of the work?

In past two decades, (3+1)-dimensional spatiotemporal optical solitons, alias light bullets (LBs)~\cite{Silberberg}, i.e. wave packets localized in three spatial and one time dimensions during propagation, have been intensively investigated due to their rich
nonlinear physics and important applications~\cite{kiv}. However, LBs studied up to now are usually produced in passive optical media, in which far-off resonance excitation schemes are used to avoid optical absorption. Such LBs have several disadvantages in
applications. For instances, the generation power of LBs in passive optical media is very high, and it is very hard to realize an active manipulation and control on them. In addition, due to off-resonance character the propagating velocity of such LBs is close to $c$ (i.e. the light speed in vacuum).

However, the disadvantages mentioned above can be overcome by using an active excitation scheme with electromagnetically induced transparency (EIT)~\cite{Imamoglu2005}. The basic principle of EIT is the use of the quantum interference effect induced by a control laser field to significantly eliminate the absorption of a probe laser field in a resonant atomic system. By using EIT, one can also realize ultraslow group velocity and giant enhancement of Kerr nonlinearity. Based on these intriguing properties, ultraslow LBs have been recently predicted in highly resonant atomic systems via EIT~\cite{LHJ}.

Particles with nonzero magnetic moments will deflect along different trajectories when passing through a gradient magnetic field. Such phenomenon, called Stern-Gerlach (SG) effect, was first discovered in the early period of quantum mechanics. As one of canonical experiments in modern physics~\cite{Ha}, SG effect is not only important for illustrating the basic concepts of quantization, spin, quantum entanglement and measurement~\cite{Sakurai,nil}, but also becomes a powerful experimental techniques in the study of molecular radicals~\cite{Kue,Ged,LBS}, metal clusters~\cite{Kni,Pok,Xu,Pay}, and nanoparticles~\cite{Per}, etc.

In a remarkable experiment carried out by Karpa and Weitz~\cite{Weitz}, a SG deflection of
a probe laser beam was observed in a $\Lambda$-type three level atomic system via EIT. But
the deflection obtained in this experiment can not be simply explained as a standard SG effect since only one ``spin'' component is involved. In addition, diffraction and dispersion inherent in the resonant  atomic system also bring a noticeable distortion of the deflected probe beam. In a recent work~\cite{Hang}, a double EIT scheme with M-type level configuration was proposed to demonstrate a SG effect of vector optical solitons, which has two polarization components (i.e. a quasispin) and allows a stable propagation of probe pulses.

However, the result in Ref.~\cite{Hang} can not be analogous to general case of SG effect in atomic physics, where space quantization of magnetic moments may result in three- and even multi-component deflection of atomic trajectories if angular-momentum  quantum number of the atoms  $J\neq 1/2$~\cite{Her}.  Such ``SG deflection spectrum''  has been widely observed in experiments and  nowadays taken to study many physical properties such as magnetic moments and spin relaxation, etc~\cite{Kue,Ged,Kni,Pok,Xu,Pay,Per}.

In this article, we propose a scheme to exhibit a SG effect of $n$-component
($n>2$) ultraslow LBs in a coherent atomic system with $(n+1)$-pod level configuration via EIT. Based on Maxwell-Bloch (MB) equations, we derive coupled (3+1)-dimensional nonlinear
Schr\"{o}dinger (NLS) equations, which govern the spatial-temporal evolution of $n$ probe-field envelopes. We show that under EIT condition a significant deflection of the $n$ components of the ultraslow LBs can be achieved by using a SG gradient magnetic field. The stability of the ultraslow LBs can be realized by an optical lattice potential contributed from a far-detuned laser field. The results presented here may have potential applications in the study of optical magnetometery, light and quantum information processing, and so on.

The rest of the article is arranged as follows. The next section describes our
model and derives the Maxwell-Bloch equations. In Sec.~\ref{Sec:3}, we derive nonlinear envelope equations of three (i.e. $n=3$) probe pulses using a method of multiple scales, and obtain ultraslow LB solutions. In Sec.~\ref{Sec:4}, the SG effect of the ultraslow LBs is studied. In Sec.~\ref{Sec:5}, we investigate SG effect of $n$ component ultraslow LBs. The last section contains a summary of our main results.

%%%%%%%%%%%%%%%%%%%%%%%%%%%%%%%%%%%%%%%%%%%%%%%%%%%%%%%%%%%%%%%%%%%%%%%%%%%%%%%%%

\section{Theoretical model }{\label{Sec:2}}

%\subsection{Model}

\begin{figure}
\includegraphics[scale=0.5]{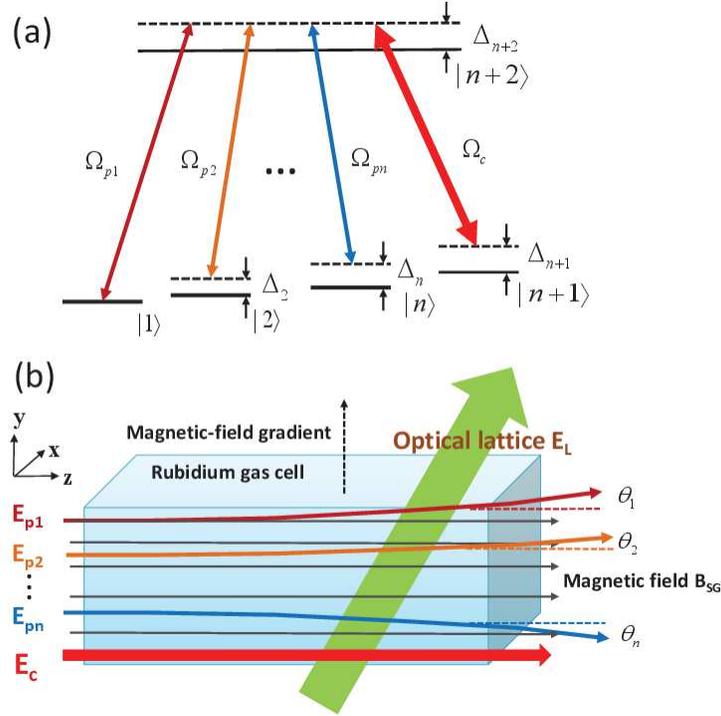}
\caption{(Color online). (a): Atomic levels and excitation scheme.
All quantities have been defined in the text.
(b): A possible arrangement of experiment for observing the SG
effect, where a SG gradient magnetic field $\textbf{B}_{\rm SG}(y)=\hat{\textbf{z}}(B_0+B_1y)$
is applied to the system.  $\theta_1,\theta_2,\ldots,\theta_{n}$ are
deflection angles of $n$ probe fields, respectively.
The (green) thick arrow denotes the far-detuned optical lattice
field $\textbf{E}_{\rm L}(x,t)=\hat{\textbf{y}}E_0\cos(x/R_\perp)\cos(\omega_{\rm L} t)$
used to stabilize LBs.  $\hat{\textbf{y}}$ and $\hat{\textbf{z}}$ are unit vectors
along $y$ and $z$ directions, respectively. The probe and control fields are co-propagating laser beams to avoid Doppler shifts.}\label{fig:1}
\end{figure}

We consider a resonant atomic system with ($n+2$) levels interacting
with $(n+1)$ laser fields. The excitation scheme constitutes a
$(n+1)$-pod configuration, where $\Omega_{pj}$ is the half Rabi
frequency of the $j$th weak, pulsed probe field ${\textbf
E}_{pj}={\textbf e}_{pj} {\cal E}_{pj}({\textbf r},t) e^{i({\textbf
k}_{pj} \cdot {\textbf r}-\omega_{pj} t)}+\textrm{c.c.}$,
$\Omega_{c}$ is the half Rabi frequency of a strong continuous-wave
control field ${\textbf E}_{c}={\textbf e}_{c} {\cal E}_{c}
e^{i({\textbf k}_{c} \cdot {\textbf r}-\omega_{c} t)}+\textrm{c.c.}$,
with ${\textbf e}_{pj}$ and ${\textbf e}_{c}$ (${\cal E}_{pj}$ and
${\cal E}_{c}$) are respectively the unit polarization vectors
(envelope functions) of the $j$th probe field and the control field,
and $\Delta_{j}$ is the detuning of the $j$th level; see
Fig.~\ref{fig:1}(a).
%
%%%%%%%%%%%%%%%%%%%%%%%%%%%%%%%%%%%%%%%%%%%%%%
%%%%%%%%%%%%%%%%%%%%%%%%%%%%%%%%%%%%%%%%%%%%%%
%
We assume that initially the atomic population is prepared in the ground states $|1\rangle$, $|2\rangle$,$\ldots$,$|n\rangle$,  and cooled to a ultracold  temperature in order to eliminate Doppler broadening and collisions. Fig.~\ref{fig:1}(b) is a possible arrangement of experimental apparatus. The aim of the co-propagating configuration of probe and control fields is also to avoid Doppler shifts.

We further assume a static SG gradient magnetic field
\begin{eqnarray} \label{B}
\textbf{B}_{\rm SG}(y)=\hat{\textbf{z}}B(y)=\hat{\textbf{z}}(B_0+B_1y)
\end{eqnarray}
is applied to the medium with $B_1\ll B_0$. Here $B_0$ contributes
to a Zeeman level shift $\Delta
E_{j,\textrm{Zeeman}}=\mu_Bg_F^{j}m_F^{j}B_0$ for level $E_j$;
$\mu_B$, $g_F^{j}$ and $m_F^{j}$ are Bohr magneton, gyromagnetic
factor, and magnetic quantum number of the level $|j\rangle$,
respectively; $B_1$ is the transverse gradient of the SG magnetic
field, which will lead to SG deflection of the probe fields.

Additionally, we assume a small, far-detuned optical lattice field
\begin{eqnarray} \label{stark}
\textbf{E}_{\rm L}(x,t)=\hat{\textbf{y}}E_0\cos(x/R_\perp)\cos(\omega_{\rm L} t)
\end{eqnarray}
is also applied into the system, where $E_0$ , $R_\perp$ , and
$\omega_{\rm L}$ are field amplitude, beam radius, and angular frequency,
respectively~\cite{note10}. Because of the existence of $\textbf{E}_{\rm L}(x,t)$, Stark
level shift $\Delta E_{j,\textrm{Stark}}=-\alpha_j\langle
E_{\rm L}^2(x,t)\rangle_t/2=-\alpha_jE_{\rm L}^2(x)/2$ occurs. Here $\alpha_j$ is the
scalar polarizability of the level $|j\rangle$, $\langle Q(x,t)\rangle_t$
represents the time average in an oscillation cycle for the quantity $Q(x,t)$,
and therefore we have $E_{\rm L}(x)=(E_0/\sqrt{2})\cos(x/R_\perp)$. The aim of introducing the
far-detuned optical field is to stabilize the LBs~\cite{Hang,Salerno}.

Under electric-dipole and rotation wave approximations, the
Hamiltonian of the atomic system in interaction picture is given by
\bea
\hat{H}_{\textrm{int}}=
& & -\sum_{j=1}^{n+2}\hbar\Delta_j|j\rangle\langle
j|-\hbar\left[\Omega_{p1}|n+2\rangle\langle 1|+\Omega_{p2}|n+2\rangle\langle
2|+\ldots+\Omega_{pn}|n+2\rangle\langle n|\right.\nonumber\\
& & +\left. \Omega_{c}|n+2\rangle\langle n+1|+\textrm{H.c.}\right].
\eea
Here $\Omega_{pj}=({\textbf e}_{pj}\cdot{\textbf p}_{j,n+2}){\cal E}_{pj}/\hbar$
and $\Omega_{c}=({\textbf e}_c\cdot {\textbf p}_{n+1,n+2}){\cal E}_c/\hbar$ are Rabi frequencies of the $j$th probe and control fields, respectively, with ${\textbf p}_{jl}$ being the electric-dipole matrix element related to the states
$|j\rangle$ and $|l\rangle$. The control field is so strong that it can be
considered to be undepleted during the propagation of probe fields.

The equation of motion for the density matrix elements in interaction picture
reads~\cite{boyd}
\begin{equation}\label{DME}
\frac{\partial \sigma}{\partial t}=-\frac{i}{\hbar}[H_{\rm
int},\sigma]-\Gamma(\sigma),
\end{equation}
where $\Gamma(\sigma)$ is a relaxation matrix denoting spontaneous
emission and dephasing. The explicit form of Eq.~(\ref{DME}) for
$n=3$ is given in Appendix {\ref{AppendixA}.

The equation of motion for probe-field Rabi frequency
$\Omega_{pj}$  can be derived by the Maxwell equation
under slowly varying envelope approximation, given by~\cite{Hua}
\begin{equation}\label{ME}
i\left(\frac{\partial}{\partial
z}+\frac{1}{c}\frac{\partial}{\partial
t}\right)\Omega_{pj}+\frac{c}{2\omega_{pj}}\left(\frac{\partial
^2}{\partial x^2}+\frac{\partial ^2}{\partial
y^2}\right)\Omega_{pj}+\kappa_{j,n+2}\sigma_{n+2,j}=0,
\end{equation}
where $\kappa_{j,n+2}={\cal N}_a\omega_{pj}|\textbf p_{j,n+2}\cdot \textbf
e_{pj}|^2/(2\epsilon_0c\hbar)$ $(j=1,2,\ldots,n)$,
with ${\cal N}_a$ being the atomic concentration.

%%%%%%%%%%%%%%%%%%%%%%%%%%%%%%%%%%%%%%%%%%%%%%%%%%%%%%%%%%%%%%%%%%%%%%%%%%%%%%%%%
\section{Nonlinear envelope equations and light bullet solutions}{\label{Sec:3}}

\subsection{Nonlinear envelope equations}{\label{Sec:31}}

One of our main aim is to get a stable propagation of all probe fields.
On the one hand, the probe fields may suffer serious distortion due to the dispersion and
diffraction of the system. On the other hand, EIT may result in a giant
enhancement of Kerr effect. It is natural to use the enhanced Kerr effect to
balance the dispersion and diffraction for obtaining solitonlike pulses
that are shape-preserved during propagation.

To this end, we use the method of multiple scales~\cite{Hua} to
derive nonlinear envelop equations of the probe fields. For
simplicity, we consider the case of $n=3$ (The case for
general $n$ will be taken into account in Sec.~\ref{Sec:5}). Taking
the asymptotic expansion $\sigma_{ml}=\sum_{\alpha=0}^{\infty}
\epsilon^{\alpha}\sigma_{ml}^{(\alpha)}$ ($m,\,l$=1-5),
$\Omega_{pj}=\sum_{\alpha=1}^{\infty}
\epsilon^{\alpha}\Omega_{pj}^{(\alpha)}$.  Here
$\sigma_{jj}^{(0)}$ is the population distribution prepared
in the state $|j\rangle$ initially, which is assumed as $1/3$ ($j=1,2,3$)
for simplicity; $\epsilon$ is a
dimensionless small parameter characterizing the typical amplitude
of the probe fields. All quantities on the right hand side of the
expansions are considered as functions of the multi-scale variables
$x_1=\epsilon x$, $y_1=\epsilon y$, $z_{\alpha}=\epsilon^{\alpha}z$
and $t_{\alpha}=\epsilon^{\alpha}t$ ($\alpha=0,2$). The SG gradient
magnetic field and the far-detuned optical lattice field are assumed
to be $B_{\rm SG}(y_1)=B_0+\epsilon^{2}B_1y_1$ and
$E_{\rm L}(x_1)=\epsilon(E_0/\sqrt{2})\cos(x_1/R_\perp)$. Thus, $\Delta_{j}$ can be expanded as
$\Delta_{j}=\Delta_{j}^{(0)}+\epsilon^{2}\Delta_{j}^{(2)}$, where
$\Delta_{2}^{(0)}=\omega_{p1}-\omega_{p2}-\omega_{21}-\mu_{21}B_0$,
$\Delta_{3}^{(0)}=\omega_{p1}-\omega_{c}-\omega_{31}-\mu_{31}B_0$,
$\Delta_{4}^{(0)}=\omega_{p1}-\omega_{41}-\mu_{41}B_0$,
$\Delta_{5}^{(0)}=\omega_{p1}+\omega_{p2}-\omega_{c}-\omega_{51}-\mu_{51}B_0$,
$\Delta_{2}^{(2)}=-\mu_{21}B_1y_1+\alpha_{21}E_0^{2}\cos^2(x_1/R_\perp)/4$,
$\Delta_{3}^{(2)}=-\mu_{31}B_1y_1+\alpha_{31}E_0^{2}\cos^2(x_1/R_\perp)/4$,
$\Delta_{4}^{(2)}=-\mu_{41}B_1y_1+\alpha_{41}E_0^{2}\cos^2(x_1/R_\perp)/4$,
and
$\Delta_{5}^{(2)}=-\mu_{51}B_1y_1+\alpha_{51}E_0^{2}\cos^2(x_1/R_\perp)/4$. Hence we have the form $d_{jl}=d_{jl}^{(0)}+\epsilon^{2}d_{jl}^{(2)}$, with $d_{jl}^{(0)}=\Delta_{j}^{(0)}-\Delta_{l}^{(0)}+i\gamma_{jl}$ and $d_{jl}^{(2)}=\Delta_{j}^{(2)}-\Delta_{l}^{(2)}$ [$\equiv-\mu_{jl}B_1y_1+\alpha_{jl}E_0^{2}\cos^2(x_1/R_\perp)/4$].

At $\alpha=1$ order, we obtain the solution in linear level
\begin{subequations}\label{eq:1st}
\begin{eqnarray}
&&\Omega_{pj}^{(1)}=F_{j}e^{i\theta_j},\\
&&\sigma_{4j}^{(1)}=-\frac{\Omega_c^{\ast}\sigma_{jj}^{(0)}}{D_j}F_{j}e^{i\theta_j},\\
&&\sigma_{5j}^{(1)}=\frac{(\omega+d_{4j}^{(0)})\sigma_{jj}^{(0)}}{D_j}F_{j}e^{i\theta_j}
\end{eqnarray}
\end{subequations}
($j=1,\,2,\,3$). Here $D_j=|\Omega_c|^2-(\omega+d_{4j}^{(0)})(\omega+d_{5j}^{(0)})$,
 $\theta_j=K_{j}(\omega)z_0-\omega t_0$, with $F_{j}$ being envelope functions
 depending on slow variables $x_1$, $y_1$, $z_2$ and $t_2$ and
 $K_j(\omega)$ being the linear dispersion relations given by
\begin{eqnarray}\label{eq:Dp}
K_{j}(\omega)=\frac{\omega}{c}+\frac{\kappa_{j5}\sigma_{jj}^{(0)}(\omega+d_{4j}^{(0)})}{D_j}.
\end{eqnarray}

In most cases $K_{j}(\omega)$ can be Taylor expanded around $\omega_{pj}$ (which corresponds to $\omega=0$)~\cite{note1}, i.e.
$K_{j}(\omega)=K_{j0}+K_{j1}\omega+\frac{1}{2}K_{j2}\omega^2+\cdots$,
with  $K_{jl}=[\partial^{l}K_{j}(\omega)/\partial\omega^{l}]|_{\omega=0}$
($l=0$, 1, 2, $\cdots$). The coefficients $K_{jl}$ have rather clear
physical interpretation, i.e. $K_{j0}={\rm Re}(K_{j0})+i{\rm
Im}(K_{j0})$ gives the phase shift per unit length and absorption
coefficient; $K_{j1}$ determines the group velocity given by
$K_{j1}=1/V_{\textrm{g}j}=1/c+\kappa_{j5}\sigma_{jj}^{(0)}(|\Omega_c|^2+(\omega+d_{4j}^{(0)})^2)/D_j^2$;
and $K_{j2}=2\kappa_{j5}\sigma_{jj}^{(0)}(|\Omega_c|^2(3\omega+2d_{4j}^{(0)}+d_{5j}^{(0)})+(\omega+d_{4j}^{(0)})^3)/D_j^3$
represent the group velocity dispersion which resulting in spreading
and attenuation of the probe pulses.

%
%%%%%%%%%%%%%%%%%%%%%%%%%%%%%%%%%%%%%%%%%%%%%%
\begin{figure}
\includegraphics[scale=0.7]{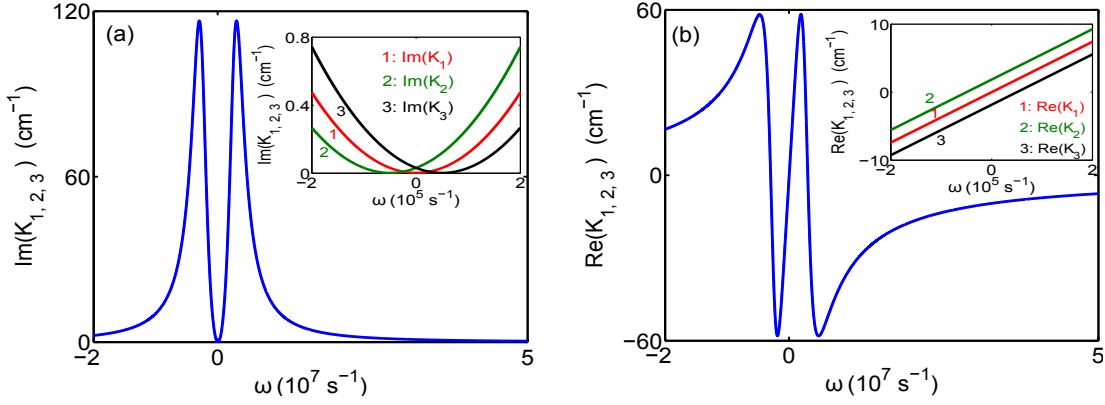}
\caption{(Color online) Linear dispersion relations of the three probe
fields. (a) $\textrm{Im}(K_j)$ and (b) $\textrm{Re}(K_j)$
($j=1,2,3$) as functions of $\omega$. Each (blue) solid curve
consists in fact of three curves, which can not be resolved since
they nearly coincide each other due to the symmetry of the system.
The inset in both panels show the absorption and dispersion curves
near $\omega=0$. The curves 1, 2, 3 in the inset of panel (a) (panel
(b)) are for Im$(K_1)$, Im$(K_2)$, Im$(K_3)$ (Re$(K_1)$, Re$(K_2)$,
Re$(K_3)$), respectively. The parameters used for plotting the
figure have been given in the text.}{\label{fig:2}}
\end{figure}
%%%%%%%%%%%%%%%%%%%%%%%%%%%%%%%%%%%%%%%%%%%%%%
%

From Eq.~(\ref{eq:Dp}) we see that the linear dispersion relation of the system has three branches. Fig.~\ref{fig:2}(a) and Fig.~\ref{fig:2}(b) show the absorption
$\textrm{Im}(K_{j})$ and dispersion $\textrm{Re}(K_{j})$
$(j=1,\,2,\,3)$ of the three probe fields as functions of $\omega$,
respectively. The parameters are chosen from a laser-cooled
$^{87}$Rb atomic gas with D$_1$ line transitions $5^{2}S_{1/2}\rightarrow5^{2}P_{1/2}$
with atomic states assigned as
$|1\rangle=|5^{2}S_{1/2}, F=2, m_F=2\rangle$,
$|2\rangle=|5^{2}S_{1/2}, F=1, m_F=0\rangle$,
$|3\rangle=|5^{2}S_{1/2}, F=1, m_F=1\rangle$,
$|4\rangle=|5^{2}S_{1/2}, F=2, m_F=1\rangle$, and
$|5\rangle=|5^{2}P_{1/2}, F=2, m_F=1\rangle$~\cite{Steck}. The decay rates are
$\gamma_{41}\approx\gamma_{42}\approx\gamma_{43}=1$ kHz, and
$\Gamma_5=5.75$ MHz. The other parameters are taken as
$\kappa_{15}\approx\kappa_{25}\approx\kappa_{35}=1.0\times10^{9}$
cm$^{-1}\cdot$ s$^{-1}$, $\Delta_{2}=-5.0\times10^4$ s$^{-1}$,
$\Delta_{3}=5.0\times10^4$ s$^{-1}$, $\Delta_{4}=\Delta_{5}=0$
s$^{-1}$, and $\Omega_{c}=3\times10^6$ s$^{-1}$. We see that
transparency windows are opened in the absorption curves Im$(K_j)$
near $\omega=0$ ($j=1,2,3$), a typical character of EIT contributed
by the control field-induced quantum interference effect. In fact,
in linear regime the excitation scheme for $n=3$ (i.e. 4-pod
configuration) consists of three independent $\Lambda$-type
three-level systems and hence possesses three dark states, each of
them displays EIT character. Note that the (blue) solid curve in
both panels is a overlapping of three curves, which can not be
resolved since they nearly coincide each other due to the symmetry
of the system. The advantage of such symmetry is very useful for
obtaining matching group velocities of different probe fields (i.e.
${\rm Re}(K_{11})\approx {\rm Re}(K_{21}) \approx {\rm Re}(K_{31})$;
Fig.~\ref{fig:2}(b)), which is very essential to obtain significant
SG deflection of the probe fields.

At $\alpha=2$ order, we obtain solvability conditions $i[\partial
F_{j}/\partial z_{1}+(\partial K_j/\partial\omega)\partial F_{j}/\partial t_{1}]=0$
($j=1,\,2,\,3$), which indicate that the envelope function
$F_j$ travels with complex group velocity $V_{{\rm g}j}\equiv(\partial K_j/\partial\omega)^{-1}$. Explicit expressions of the solution at this order have been presented in Appendix~\ref{AppendixB}.

At $\alpha=3$ order, the giant enhancement of Kerr effect of the system generated by EIT plays an important role. We obtain the coupled NLS equations governing the evolution of $F_j$:
\bea \label{NLSE}
& & i\left(\frac{\partial}{\partial
z_2}+\frac{1}{V_{\textrm{g}j}}\frac{\partial}{\partial
t_2}\right)F_{j}+\frac{c}{2\omega_{pj}} \left(\frac{\partial ^2}{\partial x_1^2}+\frac{\partial
^2}{\partial y_1^2}\right)F_{j}\nonumber\\
& & -\sum_{l=1}^{3}W_{jl}|F_{l}|^2F_{j}e^{-2\bar{a}_l z_2}
+[M_jB_1y_1+N_jE_0^2\cos^2(x_1/R_\perp)]F_{j}=0,
\eea
($j=1,\,2,\,3$). Here $\bar{a}_l=\epsilon^{-2}\textrm{Im}(K_{l0})$; $W_{jl}$ result
from the Kerr nonlinearity, which contribute to self-phase modulation (when $j=l$) and
cross-phase modulation (when $j\neq l$), with explicit expressions given in
Appendix~\ref{AppendixC}; $M_j$ and $N_j$ have the form
\begin{subequations}
\begin{eqnarray}
&&M_j=-\kappa_{j5}\frac{(\omega+d_{4j}^{(0)})^2\mu_{5j}+|\Omega_c|^2\mu_{4j}}{3D_j^2},\\
&&N_j=\kappa_{j5}\frac{(\omega+d_{4j}^{(0)})^2\alpha_{5j}+|\Omega_c|^2\alpha_{4j}}{12D_j^2},
\end{eqnarray}
\end{subequations}
characterizing the contributions of the SG gradient magnetic field and far-detunned optical
lattice field, respectively.

\subsection{Light bullet solutions}{\label{Sec:32}}

For convenience, we introduce the dimensionless variables
$s=z/L_{\textrm{Diff}}$, $\tau=t/\tau_0$,
$v_{\textrm{g}j}=V_{\textrm{g}j}\tau_0/L_{\textrm{Diff}}$, and
$u_{j}=(\Omega_{pj}/U_0)\exp[-i\textrm{Re}(K_{j0})z]$. Here
$L_{\textrm{Diff}}=\omega_pR_\perp^2/c$, $\tau_0$, and $U_0$ are
typical diffraction length, probe-field pulse duration, and half Rabi
frequency, respectively. Then Eq.~(\ref{NLSE}) can be written into
the dimensionless form
\begin{eqnarray}\label{NLSE1}
\left[i\left(\frac{\partial}{\partial
s}+\frac{1}{v_{\textrm{g}j}}\frac{\partial}{\partial
\tau}\right)+\frac{1}{2}\left(\frac{\partial ^2}{\partial
\xi^2}+\frac{\partial ^2}{\partial
\eta^2}\right)\right]u_{j}-\sum_{l=1}^{3}
\textrm{g}_{jl}|u_{l}|^2u_{j}+V_{j}(\xi,\eta)u_{j}=-iA_{j}u_{j},
\end{eqnarray}
where $\textrm{g}_{jl}=W_{jl}/|W_{12}|$ characterize nonlinear effect; and $A_j=\textrm{Im}(K_{j0})L_{\textrm{Diff}}$ ($j=$1,\,2,\,3)
are absorption coefficients. The potential functions in Eq.~(\ref{NLSE1}) read
\begin{eqnarray}\label{V}
V_j(\xi,\eta)=\mathcal{M}_j\eta+\mathcal{N}_j\cos^2(\xi),
\end{eqnarray}
with $\mathcal{M}_j=L_{\textrm{Diff}}M_jR_\perp B_1$ and
$\mathcal{N}_j=L_{\textrm{Diff}}N_jE_0^2$ ($j=1,\,2,\,3$). Note that in the
derivation of Eq.~(\ref{NLSE1}) we have assumed $\tau_0$ is large so
that group-velocity dispersion term (i.e., the term proportional to
$\partial^2 u_j /\partial \tau^2$) can be neglected, which can be
easily realized experimentally. Furthermore, the absorption can also
be negligible by choosing suitable system parameters under the
condition of EIT. In fact, when we take $\tau_0=9.0\times10^{-7}$ s,
$U_0=3.38\times10^{6}\,$s$^{-1}$,
$\Delta_2=1.2\times10^4\,$s$^{-1}$,
$\Delta_3=1.5\times10^4\,$s$^{-1}$,
$\Delta_4=2.0\times10^5\,$s$^{-1}$,
$\Delta_5=9.0\times10^6\,$s$^{-1}$,
$\Omega_c=2.0\times10^7$s$^{-1}$, and $R_\perp=36\,\mu$m with other parameters the same as in Fig.~\ref{fig:2},  we obtain the typical diffraction length $L_{\textrm{Diff}}=1.0$ cm, which is approximately equal to the typical nonlinearity length $L_{\textrm{N}}[\equiv1/(U_0^2|W_{12}|)]$. However, the typical linear absorption length $L_{\textrm{A}\,j}=1/\textrm{Im}(K_{j0})$ is around $924.0$ cm and typical second-order dispersion length $L_{\textrm{Disp}\,j}=\tau_0^2/{\textrm {Re}}(K_{j2})$ ($j=1,2,3$) is $20.5$ cm, both of them are much larger than $L_{\textrm{Diff}}$ and $L_{\textrm{N}}$. Base on the results, we thus have the ratio coefficients $d_j=L_{\textrm{Diff}}/L_{\textrm{Disp} j}\approx0.0495$, and $A_j=L_{\textrm{Diff}}/L_{\textrm{A}\,j}\approx0.0011$, which describe the significance of the various characteristic interaction lengths relative to the diffraction effect. Therefore we can safely neglect the corresponding terms because they are much smaller than 1.

With the above parameters we obtain group velocities of three probe field envelopes (i.e. $\tilde{V}_{\textrm{g}j}={\rm Re}(\partial K_j/\partial \omega)^{-1}$\,)
\bea
& & \tilde{V}_{\textrm{g}1}=3.964\times10^{-5}\,c,\\
& & \tilde{V}_{\textrm{g}2}=3.966\times10^{-5}\,c, \\
& & \tilde{V}_{\textrm{g}3}=3.967\times10^{-5}\,c. \eea We see that three
group velocities of probe field envelopes are very small comparing with $c$ ($c$ is light speed in vacuum), and nearly matched each other. The ultraslow and matched group velocities are essential to obtain significant SG deflection of the probe fields, as will be shown below.

We now seek approximated analytical solutions of the Eq.~(\ref{NLSE1})
with the form~\cite{guo}
$u_j(\rho_j,\tau,\xi,\eta)=f_j(\rho_j)v_j(\tau,\xi,\eta)$, where
$f_j(\rho_j)$ are normalized Gaussian functions, that is,
$f_j=[1/(\rho_0\sqrt{\pi})]^{1/2}\exp[-\rho_j^2/(2\rho_0^2)]$
with $\rho_j=s-v_{gj}\tau$ and $\rho_0$ a constant. Integrating out
the variable $\rho_j$, Eq.~(\ref{NLSE1}) becomes
\begin{eqnarray}\label{NLSE2}
\left[\frac{i}{v_{\textrm{g}j}}\frac{\partial}{\partial
\tau}+\frac{1}{2}\left(\frac{\partial ^2}{\partial
\xi^2}+\frac{\partial ^2}{\partial
\eta^2}\right)\right]v_{j}-\frac{1}{\sqrt{2\pi}\rho_0}\sum_{l=1}^{3}
\textrm{g}_{jl}|v_{l}|^2v_{j}+V_{j}(\xi,\eta)v_{j}=0.
\end{eqnarray}

To obtain the LB solutions of Eq.~(\ref{NLSE2}), we consider
several reasonable approximations: (i)In the presence of the SG
gradient magnetic field, the three probe-field envelopes will
separate each other after propagating some distance. In such
situation, the interaction between different envelopes becomes weak
and hence the cross-phase-modulation terms can be neglected; (ii)The
potential wells of the optical lattice are assumed to be deep enough, so that the
probe-field envelopes are almost trapped in the wells in $x$ direction.
Hence $V_j(\xi,\eta)$ given in Eq.~(\ref{V}) can be approximated by
$\mathcal{M}_j\eta+\mathcal{N}_j-\mathcal{N}_j\xi^2$. Thus
Eq.~(\ref{NLSE2}) can be approximated as
\begin{eqnarray}\label{NLSE3}
&&\left[\frac{i}{v_{\textrm{g}j}}\frac{\partial}{\partial
\tau}+\frac{1}{2}\left(\frac{\partial ^2}{\partial
\xi^2}+\frac{\partial ^2}{\partial
\eta^2}\right)\right]v_{j}-\frac{1}{\sqrt{2\pi}\rho_0}
\textrm{g}_{jj}|v_{j}|^2v_{j}+(\mathcal{M}_j\eta+\mathcal{N}_j-\mathcal{N}_j\xi^2)v_{j}=0.
\end{eqnarray}

Assuming
$v_j({\tau,\xi,\eta})=w_j(\tau,\eta)\phi_j(\xi)\exp(i\mathcal{N}_jv_{\textrm{g}j}\tau)$,
where $\phi_j(\xi)$ is a normalized ground state solution satisfying
the eigenvalue problem $(\partial ^2/\partial
\xi^2-2\mathcal{N}_j\xi^2)\phi_j=2E_{\xi}\phi_j$ with
$E_{\xi}=-\sqrt{\mathcal{N}_j/2}$, and integrating over the variable $\xi$, one obtains
\begin{eqnarray}\label{NLSE4}
&&\left(\frac{i}{v_{\textrm{g}j}}\frac{\partial}{\partial
\tau}+\frac{1}{2}\frac{\partial ^2}{\partial
\eta^2}\right)w_{j}-\frac{\mathcal{N}_j^{1/4}}{2^{3/4}\pi\rho_0}
\textrm{g}_{jj}|w_{j}|^2w_{j}+\left(\mathcal{M}_j\eta-\sqrt{\frac{\mathcal{N}_j}{2}}\right)w_{j}=0.
\end{eqnarray}
Eq.~(\ref{NLSE4}) is a (1+1)-dimensional NLS equation with a linear
potential, which admits the exact single-soliton
solutions~\cite{Yan}
\begin{eqnarray}
w_j=\mathcal{A}_je^{i\varphi_j}{\textrm {sech}}\Theta_j,
\end{eqnarray}
where $\mathcal{A}_j=(2^{5/4}\mathcal{N}_j^{1/4}\pi\rho_0/\vert{
\textrm{g}_{jj}}\vert)^{1/2}$, $\varphi_j=\mathcal{M}_jv_{{\textrm
g}j}\tau(\eta-\mathcal{M}_jv_{{\textrm g}j}^2\tau^2/6)$, and
$\Theta_j=(2\mathcal{N}_j)^{1/4}(\eta-\mathcal{M}_jv_{{\textrm
g}j}^2\tau^2/2)$. Finally, we obtain the solution of Eq.~(\ref{NLSE2})
\begin{eqnarray}\label{soliton}
u_j=\mathcal{A}_j[1/(\rho_0\sqrt{\pi})]^{1/2}(\sqrt{2\mathcal{N}_j}
/\pi)^{1/4}e^{i\varphi_j}e^{-(s-v_{{\textrm
g}j}\tau)^2/(2\rho_0^2)}e^{-\sqrt{\mathcal{N}_j}\xi^2/\sqrt{2}}{\textrm
{sech}}\Theta_j,
\end{eqnarray}
which is a nonlinear solution localized in three space and one time dimensions,
i.e. the (3+1)-dimensional LB solution of the system.

%%%%%%%%%%%%%%%%%%%%%%%%%%%%%%%%%%%%%%%%%%%%%%%%%%%%%%%%%%%%%%%%%%%%%%%%%%%%%%%%%

\section{Stern-Gerlach deflection of 3-component ultraslow light bullets}{\label{Sec:4}}

When returning to original variables, the LB solution (\ref{soliton}) has the form
\bea
\Omega_{pj}=
& & U_0 \mathcal{A}_j \left(\frac{1}{\rho_0\sqrt{\pi}}\right)^{1/2}
\left(\frac{\sqrt{2\mathcal{N}_j}}{\pi}\right)^{1/4}  e^{i\varphi_j}
e^{-\sqrt{\mathcal{N}_j} x^2/(\sqrt{2}R_{\perp}^2)}
e^{-(z-\tilde{V}_{gj}t)^2/(2L_{\rm Diff}^2\rho_0^2)} \nonumber\\
& & \times {\rm sech} \left\{\frac{(2\mathcal{N}_j)^{1/4}}{R_{\perp}}
    \left(y-\frac{\mathcal{M}_j R_{\perp} \tilde{V}_{gj}^2}{2L_{\rm Diff}^2}
     t^2 \right)\right\}. \label{LBO}
\eea
We see that the LB travels in $z$ direction with ultraslow group velocity $\tilde{V}_{gj}$.
In addition, it has an acceleration $\mathcal{M}_j R_{\perp} \tilde{V}_{gj}^2/L_{\rm Diff}^2$ in $y$
direction, which results in the SG deflection. Note that $\mathcal{M}_j$ is proportional to the
parameter $B_1$, i.e. the deflection comes from the SG gradient magnetic field given by Eq.~(\ref{B}).

Shown in Fig.~\ref{fig:3}
%
%%%%%%%%%%%%%%%%%%%%%%%%%%%%%%%%%%%%%%%%%%%%%%
%
\begin{figure}
\includegraphics[scale=0.7]{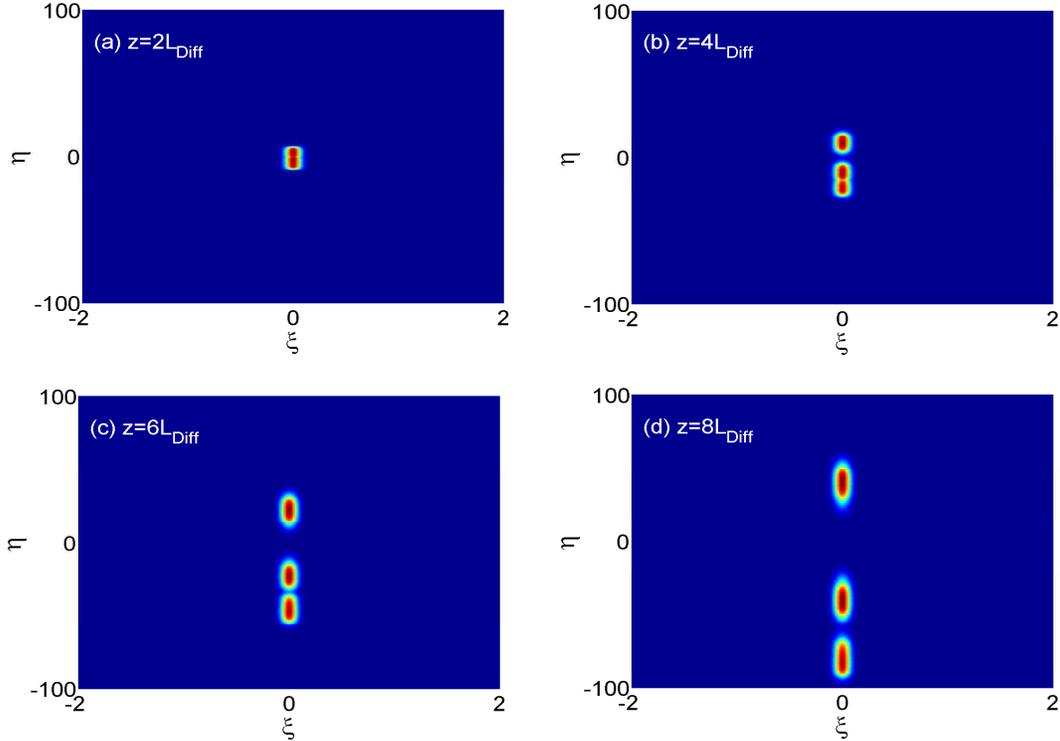}
\caption{(Color online) The SG deflection spectrum of the 3-component
ultraslow LB. (a), (b), (c) and (d) are deflections of the LB
when propagating to $z=2L_{\rm Diff}$,
$z=4L_{\rm Diff}$, $z=6L_{\rm Diff}$, and $z=8L_{\rm Diff}$,
respectively. The bright spots from top to bottom are
distributions of $|\mathcal{E}_{p1}|^2$,
$|\mathcal{E}_{p2}|^2$, and $|\mathcal{E}_{p3}|^2$ in $x$-$y$ plane,
respectively.}{\label{fig:3}}
\end{figure}
%
%%%%%%%%%%%%%%%%%%%%%%%%%%%%%%%%%%%%%%%%%%%%%%
%
is the result of the SG deflection spectrum of the 3-component LB by numerically simulating Eq.~(\ref{NLSE3}) with $B_1\neq0$. Panels (a), (b), (c), and (d) give the light intensity of the LB when propagating respectively to $z=2L_{\rm Diff}$, $z=4L_{\rm Diff}$, $z=6L_{\rm Diff}$, and $z=8L_{\rm Diff}$ for $B_1=1.2\,$mG/mm and $E_0=5.0\times10^4\,$V/m. The bright spots from top to bottom in each panel are distributions of $|\mathcal{E}_{p1}|^2$, $|\mathcal{E}_{p2}|^2$, and $|\mathcal{E}_{p3}|^2$ in $x$-$y$ plane, respectively. Through the information of Fig.~\ref{fig:3}, we can see that an obvious deflection track of LBs occurs due to the SG gradient magnetic field existing. The phenomenon is similar to the SG deflection for atoms.

We now determine the SG deflection angles of each LB components. From the solution (\ref{LBO})
we get the propagating velocity of the $j$th LB component at time $t$
\be
{\bf V}_j=\left(0, \frac{\mathcal{M}_j\tilde{V}_{{\rm g}j}^2R_\perp}{L_{\textrm{Diff}}^2} t,\tilde{V}_{{\rm g}j}\right).
\ee
Assume the medium length in the $z$ direction is $L$. The running
time in $z$ direction is thus $L/\tilde{V}_{{\rm g}j}$. At the exit of
the medium the velocity of the $j$th LB component will be ${\bf
V}_j=(0, V_{yj},\tilde{V}_{{\rm g}j})$ with $V_{yj}=\mathcal{M}_j \tilde{V}_{{\rm g}j}
R_{\perp} L/L_{\textrm{Diff}}^2$. As a result, we have the
deflection angle after passing through the medium
\begin{eqnarray}\label{theta}
\theta_j=\frac{V_{yj}}{\tilde{V}_{{\textrm g}j}}=\frac{L}{\tilde{V}_{{\textrm
g}j}}\frac{\mu_{\textrm{sol}\,j}}{p_j}r^2B_1,
\end{eqnarray}
where $r=R_\perp/L_{\textrm{Diff}}$, $p_j=\hbar k_{pj}$ is photon
momentum, and $\mu_{\textrm{sol}\,j}=L_{\textrm{Diff}}M_j
\tilde{V}_{{\textrm g}j}\hbar k_{pj}$ is {\it effective} magnetic
moment~\cite{note2}. With the data given in Fig.~\ref{fig:3} we
obtain $\mu_{\textrm{sol}\,1,\,2}=\pm3.70\times10^{-19}$ J/T and
$\mu_{\textrm{sol}\,3}=-7.41\times10^{-19}$ J/T. The center position
of the $j$th probe envelope at the exit of the medium reads
\be  \label{CP}
(x_j,y_j,z_j)=\left(0,\frac{\mathcal{M}_jL^2R_\perp}{2L_{\textrm{Diff}}^2},L\right).
\ee
From the formula~(\ref{theta}) we see that the deflection angle $\theta_j$ is inversely proportional to $\tilde{V}_{{\rm g}j}$. So the ultraslow group velocity $\tilde{V}_{{\rm g}j}$ induced by the EIT effect can result in large SG  deflection angles of the LB.

%%%%%%%%%%%%%%%%%%%%%%%%%%%%%%%%%%%%%%%%%%%%%%
\begin{figure}
\includegraphics[scale=0.75]{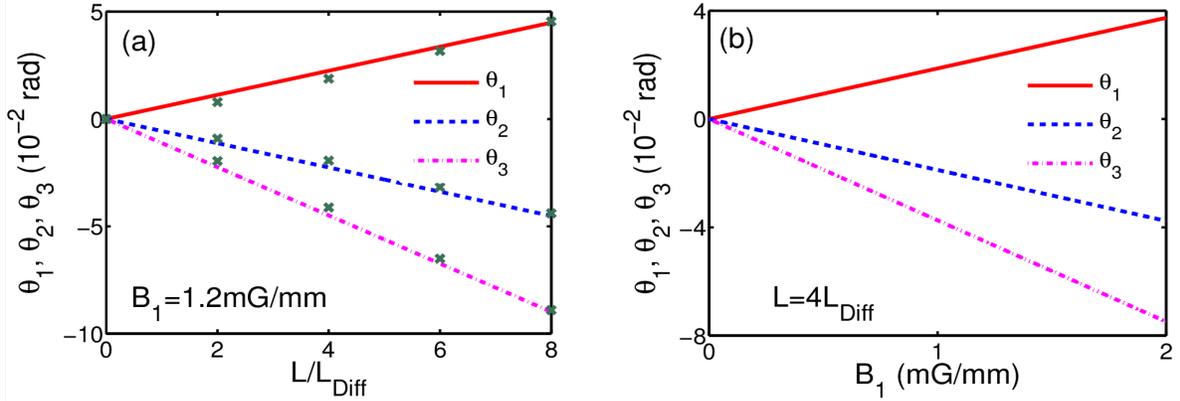}
\caption{(Color online) (a): Deflection angles of the LB components as functions of medium length $L$ for magnetic field gradient $B_1=1.2\,$mG/mm. The solid, dashed and dashed-dotted lines denote deflection angles $\theta_1$, $\theta_2$ and $\theta_3$, respectively. Points labeled by $``\times"$ are center positions of the LBs components obtained numerically.  (b): Deflection angles of the LB components as functions of $B_1$ for $L=4L_{\textrm{Diff}}$.  The results of $\theta_1$, $\theta_2$ and $\theta_3$ are respectively labeled by solid, dashed and dashed-dotted lines.}{\label{fig:4}}
\end{figure}
%%%%%%%%%%%%%%%%%%%%%%%%%%%%%%%%%%%%%%%%%%%%%%

Shown in Fig.~\ref{fig:4}(a) are deflection angles of the LB components as functions of medium length $L$ for magnetic field gradient $B_1=1.2\,$mG/mm. The solid, dashed and dashed-dotted lines denote respectively deflection angles $\theta_1$, $\theta_2$ and $\theta_3$ obtained by using the formula~(\ref{theta}) for $\tilde{V}_{\rm g1}\simeq \tilde{V}_{\rm g2}\simeq \tilde{V}_{\rm g3}\approx4.0\times10^{-5}c$. Points labeled by $``\times"$ are numerical results of the center position of LBs components obtained in Fig.~\ref{fig:3}. From the figure, we obtain $(\theta_1,\,\theta_2,\,\theta_3)=(2.25,\,-2.24,\,-4.49)\times10^{-2}$ rad for $L=4L_{\textrm{Diff}}$, which is three orders of magnitude larger than that for linear polariton obtained in \cite{Weitz}.

The SG effect of the LBs demonstrated above may show many intriguing applications. For instance, through measuring the deflection angles of LBs components, one can obtain the gradient magnetic field $B_1$. Fig.~\ref{fig:4}(b) shows the deflection angles as functions of $B_1$ for $L=4L_{\textrm{Diff}}$. The results of $\theta_1$, $\theta_2$ and $\theta_3$ are respectively labeled by solid, dashed and dashed-dotted lines. We see that the larger the magnetic field gradient, the larger the SG deflection angles. We expect that the significant SG deflection obtained here may have potential applications in optical magnetometery and quantum information processing, etc.

%%%%%%%%%%%%%%%%%%%%%%%%%%%%%%%%%%%%%%%%%%%%%%%%%%%%%%%%%%%%%%%%%%%%%%%%%%%%%%%%%
\section{SG effect of $n$-component ultraslow light bullets}{\label{Sec:5}}

We now investigate the SG deflection of multi-component ultraslow
LBs in a $(n+2)$-level system via EIT.  For the $(n+1)$-pod level
configuration ($n>3$) shown in Fig.~\ref{fig:1}(a), the theoretical
approach is a direct generation of that developed in last two
sections. Using the weak nonlinear perturbation theory~\cite{Hua},
we can obtain the following coupled NLS equations
\begin{eqnarray}\label{NLSE5}
&&i\left(\frac{\partial}{\partial
z_2}+\frac{1}{V_{\textrm{g}j}}\frac{\partial}{\partial
t_2}\right)F_{j}+\frac{c}{2\omega_{pj}}\left(\frac{\partial^2}{\partial
x_1^2}+\frac{\partial^2}{\partial
y_1^2}\right)F_{j}\nonumber\\
&&-\sum_{l=1}^{n}W_{jl}|F_{l}|^2
e^{-2{\bar a}_lz_2}F_{j}+[M_jB_1y_1+N_jE_0^2\cos^2(x_1/R_\perp)]F_{j}=0,
\end{eqnarray}
($j=1,2,\ldots,n$). Here $F_j$ is the envelope of the $j$th probe field;
$W_{jl}$ are coefficients of self-phase (for $j=l$) and
cross-phase (for $j\neq l$) modulations; $M_j$ ($N_j$) characterize the
amplitude of the SG gradient magnetic field (the far-detuned optical
lattice field). Explicit expressions of these coefficients are lengthy and
omitted here.

A similar approach as in the last two sections can also be done. For SG deflection
problem, what we want is the result in far-field approximation, i.e. the one for a larger propagation distance where different the LB components are separated away. Thus the cross-phase modulation terms in Eq.~(\ref{NLSE5}) can be neglected reasonably. By assuming a strong confinement from the far-detuned optical lattice potential, we can obtain an equation similar to (\ref{NLSE3}), and LB solution with the same form of (\ref{LBO}).

In this case, we also numerically simulate Eq.~(\ref{NLSE3}) with $B_1\neq0$ to investigate the deflection of $n$-component LBs. Shown in Fig.~\ref{fig:5} is the SG deflection spectrum of a $5$-component ultraslow LB. Panels (a), (b), (c) and (d) show the light intensity of the LB components when propagating to $z=2L_{\rm Diff}$, $z=4L_{\rm Diff}$, $z=6L_{\rm Diff}$, and $z=8L_{\rm Diff}$, respectively. We see that the LB is robust during propagation and its components separate away fast in $y$ direction. Such result can be taken as a satisfactory analog of general SG effect of atoms for angular-momentum quantum number $J>1/2$.
\begin{figure}
\includegraphics[scale=0.7]{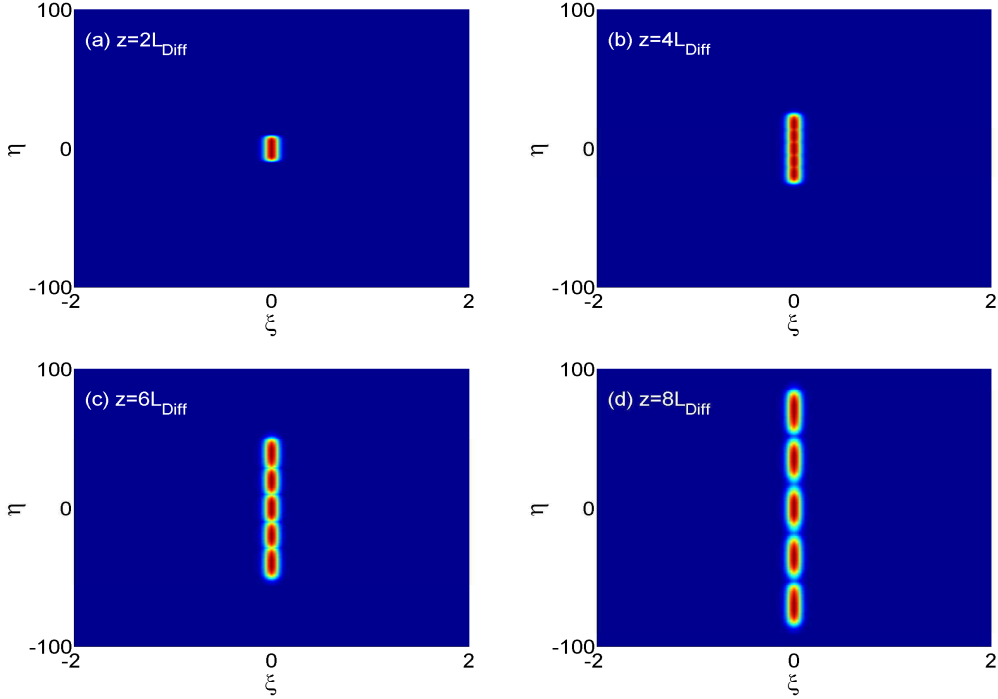}
\caption{(Color online) The SG deflection spectrum of $5$-component
ultraslow LB. (a), (b), (c) and (d) show the
deflections of the LB components when propagating to $z=2L_{\rm Diff}$, $z=4L_{\rm
Diff}$, $z=6L_{\rm Diff}$, and $z=8L_{\rm Diff}$,
respectively.}{\label{fig:5}}
\end{figure}
%

%%%%%%%%%%%%%%%%%%%%%%%%%%%%%%%%%%%%%%%%%%%%%%%%%%%%%%%%%%%%%%%%%%%%%%%%%%%%%%%%%

\section{Summary}{\label{Sec:6}}

We have suggested a scheme to exhibit a Stern-Gerlach effect of $n$-component ($n>2$)
high-dimensional ultraslow optical solitons in a coherent atomic system with
$(n+1)$-pod level configuration via EIT. Based on the MB equations, we have derived the coupled (3+1)-dimensional NLS equations governing the spatial-temporal evolution of the $n$ probe-field envelopes. We have demonstrated that under the EIT condition significant deflections of the $n$ components of the ultraslow LBs can be achieved by using a Stern-Gerlach gradient magnetic field. The stability of the ultraslow LBs can be realized by an optical lattice potential contributed from a far-detuned laser field. We expect that the results predicted here may have potential applications in the research fields of optical magnetometery, light information processing, and so on.

%%%%%%%%%%%%%%%%%%%%%%%%%%%%%%%%%%%%%%%%%%%%%%%%%%%%%%%%%%%%%%%%%%%%%%%%%%%%%%%%%

\section*{Acknowledgments}

The authors thank Chao Hang for useful discussions. This work was supported by the NSF-China under Grant Numbers
Nos.~10874043 and 11174080.

%%%%%%%%%%%%%%%%%%%%%%%%%%%%%%%%%%%%%%%%%%%%%%%%%%%%%%%%%%%%%%%%%%%%%%%%%%%%%%%%%
\appendix

\section{Equations of motion of density matrix elements}\label{AppendixA}
The elements of density matrix for $n=3$ read
\begin{subequations} \label{dme}
\begin{align}
&i\frac{\partial}{\partial
t}\sigma_{11}-i\Gamma_{15}\sigma_{55}+\Omega_{p1}^{\ast}\sigma_{51}-\Omega_{p1}\sigma_{51}^{\ast}=0,\\
&i\frac{\partial}{\partial
t}\sigma_{22}-i\Gamma_{25}\sigma_{55}+\Omega_{p2}^{\ast}\sigma_{52}-\Omega_{p2}\sigma_{52}^{\ast}=0,\\
&i\frac{\partial}{\partial
t}\sigma_{33}-i\Gamma_{35}\sigma_{55}+\Omega_{p3}^{\ast}\sigma_{53}-\Omega_{p3}\sigma_{53}^{\ast}=0,\\
&i\frac{\partial}{\partial
t}\sigma_{44}-i\Gamma_{45}\sigma_{55}+\Omega_{c}^{\ast}\sigma_{54}-\Omega_{c}\sigma_{54}^{\ast}=0,\\
&i(\frac{\partial}{\partial
t}+\Gamma_{5})\sigma_{55}+\Omega_{p1}\sigma_{51}^{\ast}+\Omega_{p2}\sigma_{52}^{\ast}+\Omega_{p3}\sigma_{53}^{\ast}\nonumber\\
&+\Omega_{c}\sigma_{54}^{\ast}-\Omega_{p1}^{\ast}\sigma_{51}-\Omega_{p2}^{\ast}\sigma_{52}-\Omega_{p3}^{\ast}\sigma_{53}-\Omega_{c}^{\ast}\sigma_{54}=0,\\
&(i\frac{\partial}{\partial
t}+d_{21})\sigma_{21}+\Omega_{p2}^{\ast}\sigma_{51}-\Omega_{p1}\sigma_{52}^{\ast}=0,\\
&(i\frac{\partial}{\partial
t}+d_{31})\sigma_{31}+\Omega_{p3}^{\ast}\sigma_{51}-\Omega_{p1}\sigma_{53}^{\ast}=0,\\
&(i\frac{\partial}{\partial
t}+d_{32})\sigma_{32}+\Omega_{p3}^{\ast}\sigma_{52}-\Omega_{p2}\sigma_{53}^{\ast}=0,\\
&(i\frac{\partial}{\partial
t}+d_{41})\sigma_{41}+\Omega_{c}^{\ast}\sigma_{51}-\Omega_{p1}\sigma_{54}^{\ast}=0,\\
&(i\frac{\partial}{\partial
t}+d_{42})\sigma_{42}+\Omega_{c}^{\ast}\sigma_{52}-\Omega_{p2}\sigma_{54}^{\ast}=0,\\
&(i\frac{\partial}{\partial
t}+d_{43})\sigma_{43}+\Omega_{c}^{\ast}\sigma_{53}-\Omega_{p3}\sigma_{54}^{\ast}=0,\\
&(i\frac{\partial}{\partial
t}+d_{51})\sigma_{51}+\Omega_{p1}(\sigma_{11}-\sigma_{55})+\Omega_{p2}\sigma_{21}
+\Omega_{p3}\sigma_{31}+\Omega_{c}\sigma_{41}=0,\\
&(i\frac{\partial}{\partial
t}+d_{52})\sigma_{52}+\Omega_{p2}(\sigma_{22}-\sigma_{55})+\Omega_{p1}\sigma_{21}^{\ast}
+\Omega_{p3}\sigma_{32}+\Omega_{c}\sigma_{42}=0,\\
&(i\frac{\partial}{\partial
t}+d_{53})\sigma_{53}+\Omega_{p3}(\sigma_{33}-\sigma_{55})+\Omega_{p1}\sigma_{31}^{\ast}
+\Omega_{p2}\sigma_{32}^{\ast}+\Omega_{c}\sigma_{43}=0,\\
&(i\frac{\partial}{\partial
t}+d_{54})\sigma_{54}+\Omega_{c}(\sigma_{44}-\sigma_{55})+\Omega_{p1}\sigma_{41}^{\ast}
+\Omega_{p2}\sigma_{42}^{\ast}+\Omega_{p3}\sigma_{43}^{\ast}=0,
\end{align}
\end{subequations}
with $d_{jl}=\Delta_{j}-\Delta_{l}+i\gamma_{jl}$. Here detunings
are defined by
$\Delta_2=\omega_{p1}-\omega_{p2}-\omega_{21}-\mu_{21}B(y)+\alpha_{21}E(x)^{2}/2$,
$\Delta_3=\omega_{p1}-\omega_{p3}-\omega_{31}-\mu_{31}B(y)+\alpha_{31}E(x)^{2}/2$,
$\Delta_4=\omega_{p1}-\omega_{c}-\omega_{41}-\mu_{41}B(y)+\alpha_{41}E(x)^{2}/2$,
$\Delta_5=\omega_{p1}-\omega_{51}-\mu_{51}B(y)+\alpha_{51}E(x)^{2}/2$, with
$\mu_{jl}=\mu_B(g_F^jm_F^j-g_F^lm_F^l)/\hbar$,
$\alpha_{jl}=(\alpha_j-\alpha_l)/\hbar$,
$\omega_{jl}=(E_j-E_l)/\hbar$. Dephasing rates are
$\gamma_{jl}=(\Gamma_j+\Gamma_l)/2+\gamma_{jl}^{\rm col}$, with
$\Gamma_j=\sum_{E_i<E_j}\Gamma_{ij}$ denoting the
spontaneous emission rates of the state $|j\rangle$ and
$\gamma_{jl}^{\rm col}$ denoting the dephasing
rate reflecting the loss of phase coherence between $|j\rangle$ and
$|l\rangle$, as might occur with elastic collisions.

\section{Explicit expressions of the second order solutions}\label{AppendixB}
The second-order solution for $n=3$ reads
\begin{subequations}\label{2ndO}
\begin{eqnarray}
&&\sigma_{21}^{(2)}=\frac{\Omega_{p1}^{(1)}\sigma_{52}^{\ast (1)}-\Omega_{p2}^{\ast (1)}\sigma_{51}^{(1)}}{\omega+d_{21}},\\
&&\sigma_{31}^{(2)}=\frac{\Omega_{p1}^{(1)}\sigma_{53}^{\ast (1)}-\Omega_{p3}^{\ast (1)}\sigma_{51}^{(1)}}{\omega+d_{31}},\\
&&\sigma_{32}^{(2)}=\frac{\Omega_{p2}^{(1)}\sigma_{53}^{\ast (1)}-\Omega_{p3}^{\ast (1)}\sigma_{52}^{(1)}}{\omega+d_{32}},\\
&&\sigma_{4j}^{(2)}=\frac{1}{D_j}\left[(\omega+d_{5j})i\frac{\partial}{\partial t_1}\sigma_{4j}^{(1)}-\Omega_c^{\ast}i\frac{\partial}{\partial t_1}\sigma_{5j}^{(1)}\right],\\
&&\sigma_{5j}^{(2)}=\frac{1}{D_j}\left[(\omega+d_{4j})i\frac{\partial}{\partial t_1}\sigma_{5j}^{(1)}-\Omega_ci\frac{\partial}{\partial t_1}\sigma_{4j}^{(1)}\right],\\
&&\sigma_{54}^{(2)}=-\frac{\Omega_{p1}^{(1)}\sigma_{41}^{\ast
(1)}+\Omega_{p2}^{(1)}\sigma_{42}^{\ast
(1)}+\Omega_{p3}^{(1)}\sigma_{43}^{\ast (1)}}{\omega+d_{54}}
\end{eqnarray}
\end{subequations}
($j=1,\,2,\,3$), where $\sigma_{41}^{(1)}$, $\sigma_{42}^{(1)}$, $\sigma_{43}^{(1)}$,
$\sigma_{51}^{(1)}$, $\sigma_{52}^{(1)}$, and $\sigma_{53}^{(1)}$
are obtained at the first order approximation.
$d_{21}=\Delta_{21}^{(0)}+i\gamma_{21}$,
$d_{31}=\Delta_{31}^{(0)}+i\gamma_{31}$,
$d_{32}=\Delta_{32}^{(0)}+i\gamma_{32}$, and
$d_{54}=\Delta_{54}^{(0)}+i\gamma_{54}$. For simplicity, the superscript of $(0)$ in $d_{jl}^{(0)}$ has been omitted in Eq.~(\ref{2ndO}).

\section{Explicit expressions of $W_{jl}$ in Eq.~(\ref{NLSE})} \label{AppendixC}
The Kerr coefficients $W_{jl}$ in  Eq.~(\ref{NLSE}) are given by
\begin{subequations}
\begin{eqnarray}
&&W_{jj}=-\frac{\kappa_{j5}|\Omega_c|^2}{3D_j^2(\omega+d_{54}^{\ast})},\,\,(j=1,\,2\,,3)\\
&&W_{12}=-\frac{\kappa_{15}}{3D_1}\left[\frac{\omega+d_{41}}{\omega+d_{21}}\left(\frac{\omega
+d_{42}^{\ast}}{D_2^{\ast}}-\frac{\omega+d_{41}}{D_1}\right)+\frac{|\Omega_c|^2}{D_2(\omega
+d_{54}^{\ast})}\right],\\
&&W_{13}=-\frac{\kappa_{15}}{3D_1}\left[\frac{\omega+d_{41}}{\omega+d_{31}}\left(\frac{\omega
+d_{43}^{\ast}}{D_3^{\ast}}-\frac{\omega+d_{41}}{D_1}\right)+\frac{|\Omega_c|^2}{D_3(\omega+d_{54}^{\ast})}\right],\\
&&W_{21}=-\frac{\kappa_{25}}{3D_2}\left[\frac{\omega+d_{42}}{\omega
+d_{21}^{\ast}}\left(\frac{\omega+d_{42}}{D_2}-\frac{\omega+d_{41}^{\ast}}{D_1^{\ast}}\right)
+\frac{|\Omega_c|^2}{D_1(\omega+d_{54}^{\ast})}\right],\\
&&W_{23}=-\frac{\kappa_{25}}{3D_2}\left[\frac{\omega+d_{42}}{\omega
+d_{32}}\left(\frac{\omega+d_{43}^{\ast}}{D_3^{\ast}}-\frac{\omega+d_{42}}{D_2}\right)
+\frac{|\Omega_c|^2}{D_3(\omega+d_{54}^{\ast})}\right],\\
&&W_{31}=-\frac{\kappa_{35}}{3D_3}\left[\frac{\omega+d_{43}}{\omega
+d_{31}^{\ast}}\left(\frac{\omega+d_{43}}{D_3}-\frac{\omega+d_{41}^{\ast}}{D_1^{\ast}}\right)
+\frac{|\Omega_c|^2}{D_1(\omega+d_{54}^{\ast})}\right],\\
&&W_{32}=-\frac{\kappa_{35}}{3D_3}\left[\frac{\omega+d_{43}}{\omega
+d_{32}^{\ast}}\left(\frac{\omega+d_{43}}{D_3}-\frac{\omega
+d_{42}^{\ast}}{D_2^{\ast}}\right)+\frac{|\Omega_c|^2}{D_2(\omega+d_{54}^{\ast})}\right],
\end{eqnarray}
\end{subequations}
where the superscript of $(0)$ in $d_{jl}^{(0)}$ has been omitted for simplicity in this equation.

%%%%%%%%%%%%%%%%%%%%%%%%%%%%%%%%%%%%%%%%%%%%%%%%%%%%%%%%%%%%%%%%%%%%%%%%%%%%%%%%%%%%

%%%%%%%%%%%%%%%%%%%%%%%%%%%%%%%%%%%%%%%%%%%%%%%%%%%%%%%%%%%%%%%%%%%%%%%%%%%%%%%%%%%%%%%%%%%%%%%%%%%%%%%%%%%%%%%%%%%%%%%%%


\begin{thebibliography}{}
\bibitem{note1} In the atomic medium, the frequency and wave number of the $j$th probe field are given by $\omega_{pj}+\omega$ and $k_{pj}+K_j(\omega)$,  respectively. Thus $\omega=0$ corresponds to the center frequency of all probe fields.
\bibitem{note2}Photons in vacuum have no magnetic momemt, and hence experience no force and have no trajectroy deflection when passing through an inhomogeneous magnetic field. However, photons may acquire effective magnetic moments in atomic media, thus experience a magnetic force and display the SG effect.
\end{thebibliography}
\end{document}